\documentclass[runningheads]{llncs}
\usepackage{graphicx}

\usepackage{hyperref}
\usepackage{color,soul}
\usepackage{listings}
\hypersetup{hidelinks,
backref=true,
pagebackref=true,
hyperindex=true,
breaklinks=true,
colorlinks=true,
urlcolor=blue,
bookmarks=true,
bookmarksopen=false,
pdftitle={Title},
pdfauthor={Author}}

\usepackage{spverbatim}

\usepackage{fancyvrb}
\usepackage{fvextra}

\usepackage{tabularx}
\usepackage{multirow, booktabs}
\usepackage[misc,geometry]{ifsym}

\usepackage{multicol}           
\usepackage[utf8]{inputenc}     
\usepackage[table]{xcolor}      

\newcommand{\beginsupplement}{%
        \setcounter{section}{0}
        \renewcommand{\thesection}{S\arabic{section}}%
        \setcounter{table}{0}
        \renewcommand{\thetable}{S\arabic{table}}%
        \setcounter{figure}{0}
        \renewcommand{\thefigure}{S\arabic{figure}}%
     }

\begin{document}
\title{Integrative Imaging Informatics for Cancer Research: Workflow Automation for Neuro-oncology (I3CR-WANO)}
%
\titlerunning{Workflow Automation for Neuro-oncology}

\author{
Satrajit Chakrabarty\inst{1}\Letter \and
Syed Amaan Abidi\inst{2} \and 
Mina Mousa\inst{2} \and 
Mahati Mokkarala\inst{2} \and 
Isabelle Hren\inst{3} \and 
Divya Yadav\inst{4} \and 
Matthew Kelsey\inst{2} \and
Pamela LaMontagne\inst{2} \and
John Wood\inst{4} \and
Michael Adams\inst{4} \and
Yuzhuo Su\inst{4} \and
Sherry Thorpe\inst{4} \and
Caroline Chung\inst{4} \and
Aristeidis Sotiras\inst{2,5} \and
Daniel S. Marcus\inst{2} 
}

\authorrunning{S. Chakrabarty et al.}
%
\institute{Department of Electrical and Systems Engineering, Washington University in St. Louis, St. Louis, MO 63130, USA \and
Mallinckrodt Institute of Radiology, Washington University School of Medicine, St. Louis, MO 63110, USA \and
Department of Computer Science and Engineering, Washington University in St. Louis, St. Louis, MO 63130, MO, USA \and
Department of Radiation Oncology, Division of Radiation Oncology, The University of Texas MD Anderson Cancer Center, 1515 Holcombe Boulevard, Houston, TX 77027, USA \and
Institute for Informatics, Washington University School of Medicine, St. Louis, MO 63110, USA \\
\Letter~Corresponding author: \email{satrajit.chakrabarty@wustl.edu}}

%
\maketitle              
\begin{abstract}
Efforts to utilize growing volumes of clinical imaging data to generate tumor evaluations continue to require significant manual data wrangling owing to the data heterogeneity. Here, we propose an artificial intelligence-based solution for the aggregation and processing of multisequence neuro-oncology MRI data to extract quantitative tumor measurements. Our end-to-end framework i) classifies MRI sequences using an ensemble classifier, ii) preprocesses the data in a reproducible manner, iii) delineates tumor tissue subtypes using convolutional neural networks, and iv) extracts diverse radiomic features. Moreover, it is robust to missing sequences and adopts an expert-in-the-loop approach, where the segmentation results may be manually refined by radiologists. Following the implementation of the framework in Docker containers, it was applied to two retrospective glioma datasets collected from the Washington University School of Medicine (WUSM; n = 384) and the M.D. Anderson Cancer Center (MDA; n = 30) comprising preoperative MRI scans from patients with pathologically confirmed gliomas. The scan-type classifier yielded an accuracy of over 99\%, correctly identifying sequences from 380/384 and 30/30 sessions from the WUSM and MDA datasets, respectively. Segmentation performance was quantified using the Dice Similarity Coefficient between the predicted and expert-refined tumor masks. Mean Dice scores were 0.882 (±0.244) and 0.977 (±0.04) for whole tumor segmentation for WUSM and MDA, respectively. This streamlined framework automatically curated, processed, and segmented raw MRI data of patients with varying grades of gliomas, enabling the curation of large-scale neuro-oncology datasets and demonstrating a high potential for integration as an assistive tool in clinical practice.

\end{abstract}

\section{Introduction}
Advancements in \textit{in vivo} neuroimaging have resulted in large volumes of clinically acquired, high-dimensional MRI datasets. Quantitative analysis of multimodal imaging data in clinical settings is essential for assessing tumor burden and treatment response in an objective and noninvasive fashion. The scientific community has sought to capitalize on recent advancements of artificial intelligence (AI)-assisted approaches in neuro-oncology~\cite{rudie2019emerging} to develop AI-driven software packages~\cite{gibson2018niftynet,beers2021deepneuro,davatzikos2018cancer} for automating neuro-oncology workflows. 

However, the distillation of high-dimensional and multimodal imaging information into meaningful quantitative information remains an ongoing pursuit~\cite{chung2021cancer}. A major challenge is that AI-driven tools typically require manually curated, preprocessed datasets comprising scans of specific sequence(s). Manual curation and preparation of imaging data are time-consuming and error-prone for multiple reasons, including non-standardized naming conventions of scans across different manufacturers or acquisition protocols~\cite{ooijen2019quality}, same series descriptions of scans irrespective of acquisition type, and missing metadata~\cite{hirsch2015we}. This challenge has been magnified by the growing number of diverse and complementary image acquisition protocols that make parsing these datasets using automated methods or heuristics extremely difficult. Nonetheless, limited efforts have been invested in automated tools for scan-type classification and data curation~\cite{remedios2018classifying,van2021deepdicomsort}. Importantly, these have not been integrated into existing software tools. These practical challenges have limited the widespread adoption of existing software packages for heterogeneous clinical imaging data.

To centralize these efforts and expedite the translation of state-of-the-art AI models from research to clinical practice, we have developed an end-to-end AI-driven framework called Integrative Imaging Informatics for Cancer Research: Workflow Automation for Neuro-oncology (I3CR-WANO), which classifies MRI sequences using an ensemble of natural language processing (NLP) and convolutional neural network (CNN) models, preprocesses the data in a reproducible manner, and segments tumor tissue subtypes using CNNs, enabling the extraction of radiomic features. Additionally, I3CR-WANO is robust to missing sequences and adopts an expert-in-the-loop approach, where the segmentation results may be manually refined by radiologists. In this study, we implemented this framework for low- and high-grade gliomas using pre-contrast T1-weighted, (T1WI), post-contrast T1WI (Gd-T1WI), T2-weighted (T2WI), and fluid-attenuated inversion recovery (FLAIR) sequences. All framework components were packaged as Docker containers to ensure reproducibility across platforms and to facilitate dissemination and deployment. For greater flexibility, the core components of the framework were implemented as independent modules instead of as a single monolithic structure.

\section{Materials and methods}
\subsection{End-to-end framework}
\begin{figure}[!htbp]
\centering
\includegraphics[width=\textwidth]{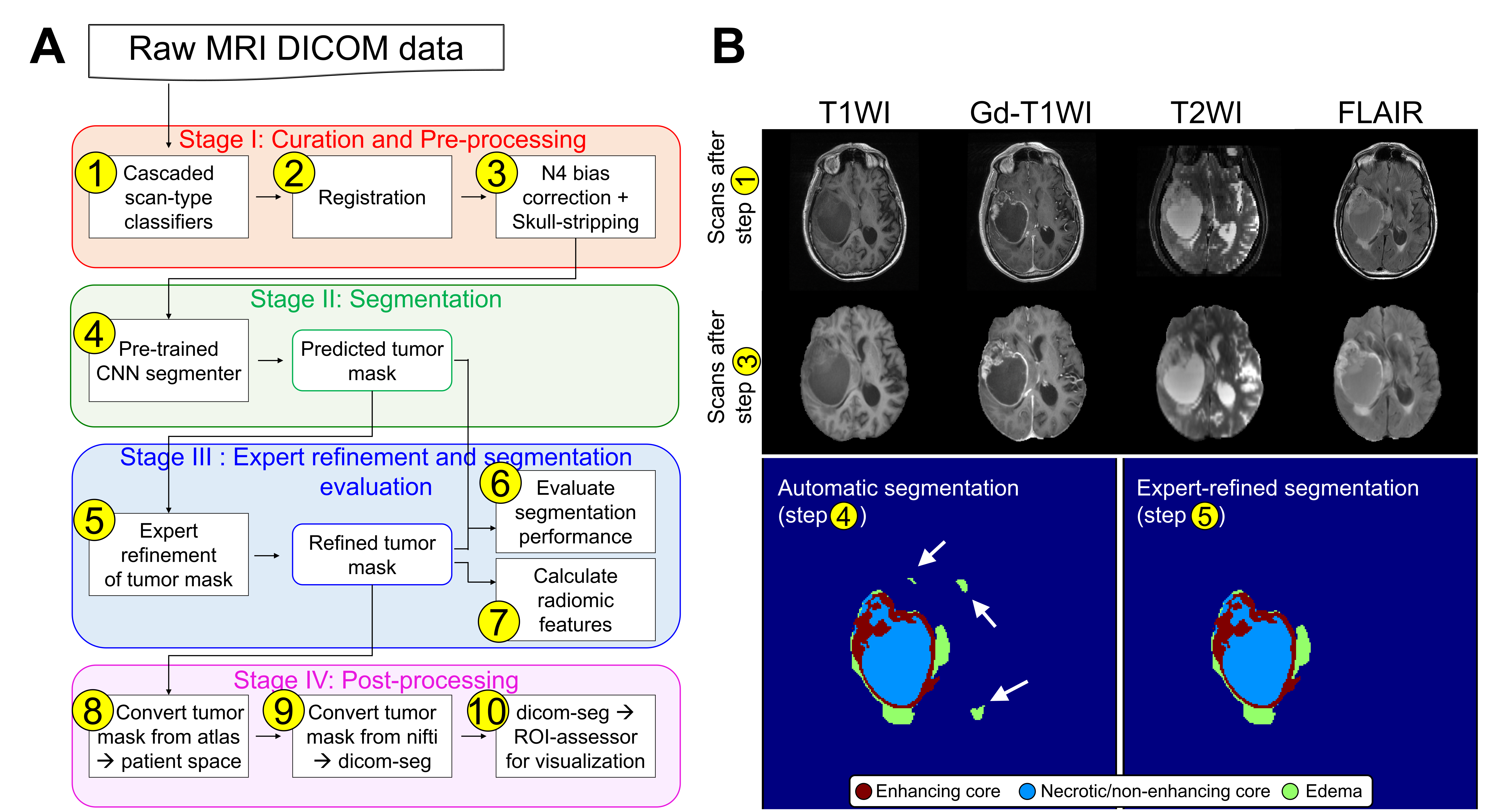}
\caption{(A) End-to-end processing framework of I3CR-WANO consisting of the following stages: I) data curation and image pre-processing, II) segmentation, III) expert refinement of tumor mask and segmentation evaluation, and IV) post-processing of the tumor mask. (B) Outputs of framework at different steps: unprocessed pre- and post-contrast T1-weighted (T1WI, Gd-T1WI), T2-weighted (T2WI), and FLAIR sequences (step 1), same sequences after pre-processing (step 3), tumor segmentation predicted by convolutional neural network (step 4), and tumor segmentation after refinement (step 5). White arrows in tumor segmentation mask image point to errors, which were corrected during refinement.} 
\label{fig:fig1}
\end{figure}

I3CR-WANO (Figure~\ref*{fig:fig1}A) consists of the following stages: I) image curation and preprocessing, II) segmentation, III) expert refinement of the tumor mask and segmentation evaluation, and IV) post-processing and visualization of the segmentation mask.

\subsubsection{Curation and pre-processing} \label{sec:curation}
The first step in the framework takes MRI data in the Digital Imaging and Communications in Medicine (DICOM) format as input and identifies the sequences that can be used in downstream segmentation and feature extraction tasks. The scan-type classifier adopts a cascade architecture. The first stage of the cascade is based on an NLP classifier (Classifier1)~\cite{chakrabarty2020preprocessing}, whereas the second stage is based on a CNN classifier (Classifier2)~\cite{van2021deepdicomsort}. For both classifiers, previously published pretrained models were used~\cite{van2021deepdicomsort,chakrabarty2020preprocessing}. Classifier1 uses information regarding the DICOM series description (i.e., tag [0008, 103E]) and the number of instances per series to classify every scan into one of two classes: segmentable or non-segmentable. Subsequently, Classifier2 performs a more granular classification of these potentially segmentable scans into T1WI, Gd-T1WI, T2WI, FLAIR, and non-segmentable classes. Additionally, the orientation of each scan (i.e., axial, coronal, sagittal) is determined by leveraging the DICOM MR acquisition type (tag [0018,0023]) and image orientation (patient) (tag [0020,0037]) information. 

In the event of multiple occurrences of T1WI, Gd-T1WI, T2WI, or FLAIR sequences, we prioritize i) axial scans over sagittal and coronal scans and ii) scans with a higher number of instances. Based on the absence of certain DICOM tags or sequence types, a particular scan or an entire session is excluded from the subsequent processing (Supplementary~\ref*{supp_methods_exclusions}). In all other cases, the curation process results in a maximum of four scans (one each for T1WI, Gd-T1WI, T2WI, and FLAIR sequences) that are used for downstream processing (Figure~\ref*{fig:fig1}B, “scans after step 1”).

I3CR-WANO comprises the following preprocessing functionalities: registration, N4-bias correction, skull-stripping, and intensity normalization. During registration, for every session, the scan with the highest number of instances among those identified in the previous step is selected as the target scan for co-registration. Then, all other scans are rigidly co-registered to the target scan, followed by affine registration to a common anatomical atlas~\cite{rohlfing2010sri24}. The transformation matrix from patient-space to atlas-space (patient2atlasmat) is stored and later used as described in Section~\ref*{sec:postproc}. For all registration steps, we used FMRIB’s Linear Image Registration Tool (FLIRT)~\cite{jenkinson2002improved,jenkinson2001global}. Next, we perform N4 bias field correction and skull-stripping of the registered sequences using the Robust Brain Extraction (ROBEX)~\cite{iglesias2011robust} tool (Figure~\ref*{fig:fig1}B, “scans after step 3”). This is followed by an image intensity normalization step, where intensities within the brain are normalized to zero mean and unit variance after excluding intensities below the 5th and above the 95th percentile. 

\subsubsection{Segmentation} \label{sec:seg}
In this step, the preprocessed scans are segmented using pretrained segmentation models (Supplementary~\ref*{supp_methods_seg_pretrain}) to produce a multiclass tumor segmentation mask (Figure~\ref*{fig:fig1}B, “automatic segmentation”). The mask comprises the edema (ED), non-enhancing/necrotic tumor core (NC), and enhancing tumor (ET) classes. Additional outputs include the tumor core (TC) class, which is created by combining the ET and NC classes, and the whole tumor (WT) class, which is constructed by combining all classes. 

To render our framework robust to missing sequences, we have trained segmentation models on different combinatorial subsets of sequences (e.g., Gd-T1WI+T2WI, Gd-T1WI+FLAIR, only Gd-T1WI, etc.). Depending on the available sequences, the segmentation module produces a multi-class segmentation mask comprising NC, ET, ED classes (if at least a Gd-T1WI is available), a binary WT segmentation mask (in the absence of Gd-T1WI but presence of T2WI and/or FLAIR), or no mask (in the absence of Gd-T1WI, T2WI, and FLAIR).

\subsubsection{Expert refinement and segmentation evaluation}
Automated segmentation models are prone to errors~\cite{baid2021rsna} like occasional erroneous labelling of vessels within the peritumoral edematous area, periventricular white matter hyperintensities, choroid plexus, and areas of Gd-T1WI bright blood products. To address this, we have adopted an expert-in-the-loop approach, where the predicted segmentation mask is optionally sent to radiologists for refinement (Figure~\ref*{fig:fig1}B, “Expert-refined segmentation”). Subsequently, the refined mask is compared with the predicted mask to evaluate the performance of the segmentation model.

\subsubsection{Post-processing} \label{sec:postproc}
Once the segmentation mask is created, the mask is warped back to the patient space by inverting the patient2atlasmat transformation matrix generated in the registration step and applying it to the tumor mask using nearest-neighbor interpolation. Subsequently, this mask is converted to a DICOM Segmentation image object format using the itkimage2segimage command from the dcmqi~\cite{herz2017dcmqi} tool. The mask can also be used for downstream processing such as the extraction of quantitative features from the tumor mask. This is supported by our framework using the PyRadiomics~\cite{van2017computational} tool, which can calculate first-order statistics, 3D shape features, and texture features for every combination of tumor class and input sequence (Supplementary~\ref*{supp_methods_radiomics}).

\subsection{Framework implementation and distribution}
The source code, detailed documentation, pre-trained AI models as well as a live demonstration of I3CR-WANO are made publicly available for non-commercial use at \url{https://github.com/satrajitgithub/NRG_AI_NeuroOnco_preproc} and \url{https://github.com/satrajitgithub/NRG_AI_NeuroOnco_segment}. In addition, to ensure portability and easy deployment, the framework has been packaged into Docker images available from DockerHub (\url{https://hub.docker.com/r/satrajit2012/nrg_ai_neuroonco_preproc} and \url{https://hub.docker.com/r/satrajit2012/nrg_ai_neuroonco_segment}).
The segmentation Docker image has been implemented using the NVIDIA GPU CLOUD runtime to leverage NVIDIA Graphics processing units during test-time inference. Besides Docker usage through the command-line interface, we also provide a visual interface of the framework through its integration with the open-source Extensible Neuroimaging Archive Toolkit (XNAT) informatics platform~\cite{marcus2007extensible}. XNAT equips the user with advanced functionalities, such as launching a Docker on a batch of sessions (batch mode) or launching a sequence of Docker commands (command orchestration). Additionally, through its integrated Open Health Imaging Foundation (OHIF) Viewer~\cite{doran2022integrating}, XNAT provides the user with a powerful set of image visualization and annotation tools, including the ability to natively refine tumor annotations, perform measurements, and save those contours and segmentation objects back into XNAT. Documentation for command-line Docker usage as well as XNAT usage are available at \url{https://github.com/satrajitgithub/NRG_AI_NeuroOnco_preproc/tree/master/documentation}.

\subsection{Application on glioma datasets}
I3CR-WANO was validated on two independent clinical datasets (Supplementary Table~\ref*{supp_table_dataset}) acquired from the retrospective health records of the Washington University School of Medicine (WUSM; n = 384; median age 56 years, range 44 – 66 years, 154 females, 230 males) and the M.D. Anderson Cancer Center (MDA; n = 30; median age 58 years, range 44 – 65 years, 15 females, 15 males). Data collected from WUSM and MDA were obtained with Institutional Review Board (IRB) approval and met the criteria for the general waiver of consent and waiver of Health Insurance Portability and Accountability Act (HIPAA) authorization.  For both datasets, the only inclusion criterion was pathologically confirmed glioma (grade II-IV) from preoperative patients with no prior resection. To ensure the broad applicability of the framework to heterogeneous clinical data, no exclusions were made based on the image acquisition parameters, image quality, or glioma grade. The segmentation models used in the framework were pre-trained on the Brain Tumor Segmentation Challenge (BraTS) 2021~\cite{baid2021rsna} dataset, publicly available from the Synapse platform (\url{https://www.synapse.org/#!Synapse:syn27046444/wiki/616992}).

\subsection{Statistical analyses}
The performance of the scan-type classifier was quantified using the overall accuracy, F1 score for each class, and confusion matrix showing the error distribution across different classes. Failures during preprocessing were identified through visual inspection. Segmentation performance was assessed using the Dice Similarity Coefficient (DSC) metric for the WT, TC, and ET classes. For the BraTS 2021 dataset, the DSC was calculated between the predicted segmentations and the provided expert-annotated ground truths. For both the WUSM and MDA datasets, the predicted tumor segmentations from the framework were refined by experts, and these expert-refined tumor masks were used as surrogate ground truths. Differences in DSC between groups with and without certain sequences were calculated using the Welch’s t-test. For all statistical tests, the threshold for statistical significance was set at $P < .05$.

\section{Results}
\begin{figure}[!htbp]
\centering
\includegraphics[width=\textwidth]{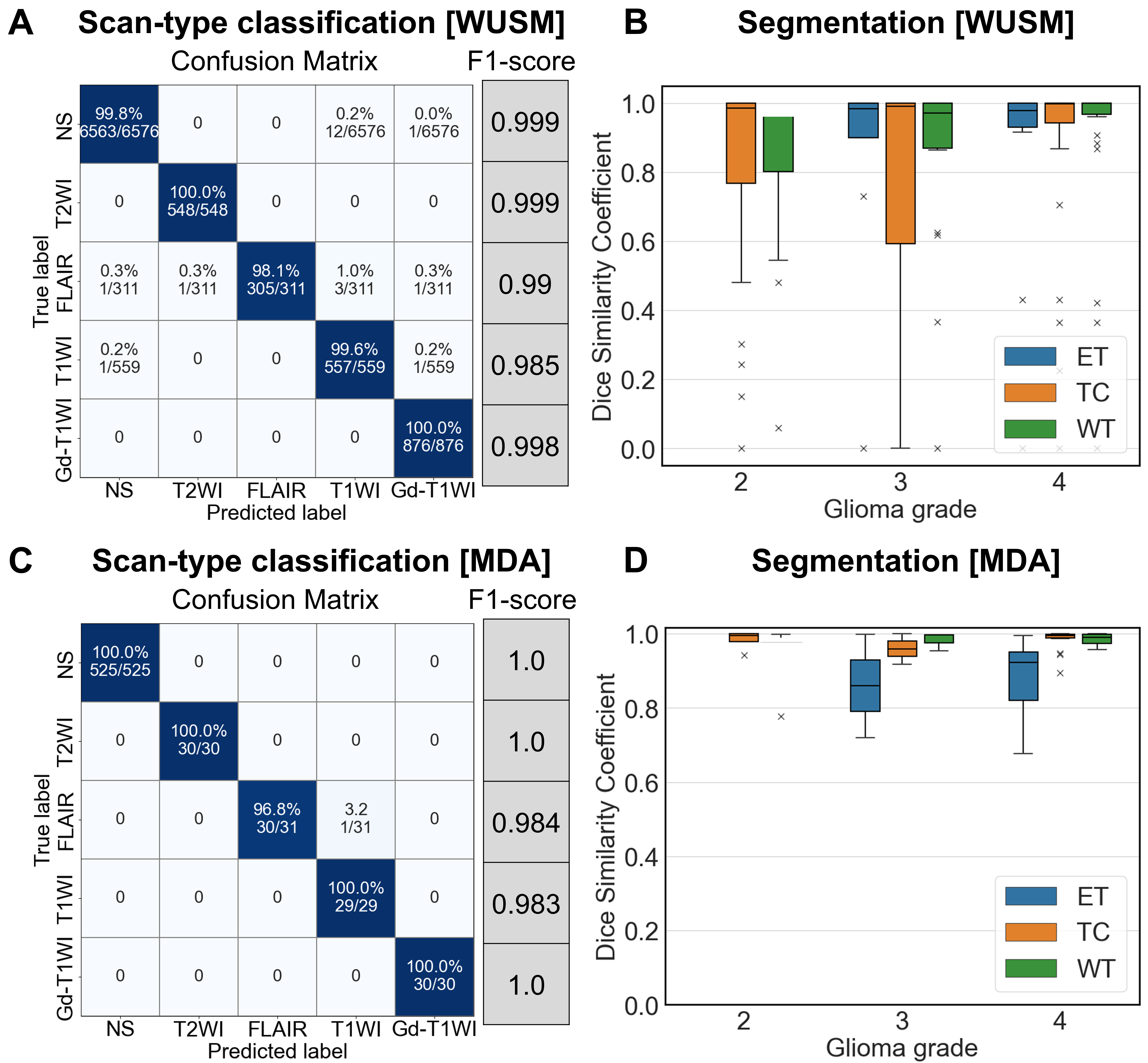}
\caption{(A), (C) Confusion matrix and F1-scores demonstrating the performance of the scan-type classifier on the WUSM and MDA datasets, respectively. (B), (D) Dispersion of Dice Similarity Coefficient values, stratified by tumor grade, showing the agreement in terms of overlap between the tumor segmentation masks predicted by convolutional neural network and the masks after refinement by a radiologist for WUSM and MDA datasets, respectively. NS = non-segmentable, WT = whole tumor, TC = tumor core, ET = enhancing tumor, WUSM = Washington University School of Medicine, MDA = M.D. Anderson Cancer Center.} 
\label{fig:fig2}
\end{figure}

\subsection{Scan-type classification and pre-processing}
The scan-type classifier yielded high overall accuracy (99.61\%, 8835/8870 scans) across the five classes (Figure~\ref*{fig:fig2}A). Precision values were above 0.99 for all classes except T1WI. The precision for T1WI (0.97, 557/572 scans) showed a minor drop because 12 non-segmentable and 3 FLAIR scans were misclassified as T1WI. Of the 12 non-segmentable scans, 10 were determined to be non-brain MR scans (i.e., spinal). In terms of recall, the classifier yielded values above 0.99 for all classes except FLAIR, which had a slightly lower recall of 0.98, because six scans were misclassified. Of the six FLAIR scans, three were misclassified as T1WI. These were all sagittal scans with a very low axial resolution. Overall, of the 384 WUSM sessions, the scan-type classifier identified all possible segmentable sequences for 380 sessions. For these 380 sessions, there were no failures in terms of preprocessing. Hence, they were used for the subsequent segmentation and radiomic feature extraction.

For the MDA dataset, the classifier yielded a very high overall accuracy (99.84\%, 643/644 scans) across five classes, with only a single sagittal FLAIR scan with low axial resolution being misclassified as T1WI. Overall, of the 30 MDA sessions, the scan-type classifier could identify all possible segmentable sequences for all sessions, and there were no failures in terms of preprocessing. Hence, all the sessions were used for subsequent segmentation and radiomic feature extraction.

\subsection{Segmentation}
The segmentation results varied depending on the input sequences, with an overall deterioration in performance with increasing number of missing sequences, especially for ET in the absence of Gd-T1WI (Supplementary~\ref*{supp_results_segmentation}). On the WUSM data, the segmentation models yielded high mean DSC for WT (0.882±0.244), TC (0.75±0.334), and ET (0.91±0.24). For all tumor classes, the mean DSCs for WHO grade IV (WT:0.901, TC:0.869, ET:0.879) were higher than those for WHO grade II (WT:0.87, TC:0.82) and grade III (WT:0.852, TC:0.778, ET:0.832) (Figure~\ref*{fig:fig2}B). 

On the MDA data, a high mean DSC was obtained for WT (0.977±0.04), TC (0.984±0.028), and ET (0.899±0.097). Overall, the model generalized well for both the external datasets.

\section{Discussion}
In this paper, we proposed I3CR-WANO, an AI-driven framework for curation, pre-processing, segmentation, and radiomic feature extraction in neuro-oncology MR studies. Through its end-to-end operation, the framework transforms unstructured DICOM MR data into quantitative 3D measurements of tumors, which can be directly used for predicting treatment response and overall survival. I3CR-WANO was validated in 414 patient cases acquired from two different clinical sites, with good overall performance in all facets of processing. The different AI models used for scan-type classification and segmentation generalized well on unseen data. The source code, dockers, and all pre-trained models of this study have been made publicly available. 

In recent years, AI-assisted tools such as niftynet~\cite{gibson2018niftynet}, DeepNeuro~\cite{beers2021deepneuro}, and the cancer imaging phenomics toolkit (CaPTk)~\cite{davatzikos2018cancer} have been proposed for automating neuro-oncology workflows in clinical practice. These tools typically depend on carefully curated data sets. However, they did not address the problem of scan-type classification or data curation within their operations. Instead, they rely on manual interaction, which is often the most time-consuming step in an AI workflow~\cite{montagnon2020deep}. In contrast, I3CR-WANO provides a more holistic solution, from data curation to radiomic feature extraction, and completely obviates the need for any intermediate manual interaction. Thus, it greatly facilitates the generation of datasets required for the development and validation of models supporting quantitative tumor measurements. Additionally, the framework’s modular structure includes the necessary commonalities in upstream pipelining that allow cascading with a wide array of downstream applications (e.g., the curation and pre-processing modules are not application-specific).

The proposed framework can streamline clinical workflows and support decision making by automating tumor segmentation and characterization. In this emerging era of precision diagnostics, the quantitative volumetric tumor measurements extracted from this framework can drive personalized treatment planning and response assessment (e.g., Response Assessment in Neuro-Oncology [RANO] criteria~\cite{van2011response}). The generated segmentation masks can be used to track tumors longitudinally and quantitatively assess their growth. In a research setting, it can significantly reduce the latency of data curation, thus expediting model prototyping and facilitating the creation of standardized large-scale neuro-oncology datasets for multi-institutional collaborations~\cite{baid2021rsna,baheti2021brain} that attempt to establish public benchmarks for various aspects of quantitative tumor analysis.

This study has certain limitations that merit discussion. First, the CNN-based scan-type classifier is currently pre-trained only on axial scans and has a minor performance drop, particularly on coronal or sagittal FLAIR scans, with a very low off-plane resolution. Second, segmentation models are currently trained on preoperative glioma cases and cannot be used on postoperative images. However, the current curation and pre-processing modules of the framework are applicable to any MRI study, irrespective of the pathology or treatment status. Moreover, owing to the modular nature of the framework, both limitations can be addressed by using more advanced containerized models that can be simply used as drop-in replacements for the current models. This flexibility can also enable the extension of this framework to multiple tumor types by integrating tumor classification models~\cite{chakrabarty2021mri} and cascading it with segmentation, radiomics, and quantitative report-generation utilities tailored for specific tumor types.

In conclusion, we developed I3CR-WANO, an AI-driven framework that transforms raw MRI DICOM data of patients with high- and low-grade gliomas to quantitative tumor measurements through systematic data curation, processing, tumor segmentation, and radiomic feature extraction, without the requirement of any manual intervention. This work can streamline clinical workflows and support clinical decision-making by automating tumor segmentation and characterization as well as help in curating large-scale neuro-oncology datasets. 

\bibliographystyle{ama}
\bibliography{mybibliography}

\section*{\hfil Supplementary Data \hfil}

\beginsupplement

\section{Supplementary Methods}
\subsection{Conditions for excluding scan or session from processing} \label{supp_methods_exclusions}
In the following conditions, a scan or an entire session is excluded from subsequent processing:
\begin{itemize}
    \item For the scan-type classification step (specifically for Classifier1), the series description tag is compulsory. So, in its absence, a particular scan is excluded from all subsequent processing.
    \item Scans that do not contain the string ‘ORIGINAL/PRIMARY’ in DICOM Image Type attribute (i.e., DICOM tag (0008,0008)) are excluded.
    \item Scans that have the DICOM Angio Flag attribute (i.e., DICOM tag (0018,0025)) set to ‘Y’ (i.e., image is Angio), are excluded. 
    \item In the absence of Gd-T1WI, T2WI, and FLAIR sequences for a particular session, the entire session is marked as not segmentable and excluded from subsequent processing.
\end{itemize}

\subsection{Pre-training of segmentation models} \label{supp_methods_seg_pretrain}
For this purpose, we adopted a 3D CNN architecture~\cite{isensee2017brain} and trained the model to produce multi-class tumor segmentation maps consisting of vasogenic edema, necrotic/non-enhancing core, and enhancing core. 
For development of the segmentation models, the BraTS 2021~\cite{baid2021rsna,menze2014multimodal,bakas2017advancing,bakas2018identifying} dataset was split into cross-validation (n = 1251) and testing (n = 219) cohorts. For hyperparameter tuning, we performed a random search~\cite{bergstra2012random} using five-fold cross-validation on the cross-validation data by using 80\% of the data for training (n =1000) and 20\% for validation (n = 251). The hyperparameters that yielded the best cross-validation results were then selected as the “best” hyperparameters. Next, the model was trained on 100\% of this data (n = 1251) using the “best” set of hyperparameters and subsequently used for prediction on the hold-out test data (n = 219).
Scans from the BraTS 2021 dataset were already registered to the SRI24 anatomical atlas~\cite{rohlfing2010sri24}, resampled to 1-mm3 isotropic resolution and skull-stripped. Additionally, for every scan, image intensities within the brain were normalized to zero mean and unit variance after excluding intensities below the 5th and above the 95th percentile. Subsequently, the image volumes were resized to a dimension of 128x128x128 before being fed into the network. We used Adam~\cite{kingma2014adam} optimizer with multi-class dice loss function. The training batch size was set to 5. The network was trained for 300 epochs with an early-stopping callback, which stopped the training after 50 epochs if the validation loss was not improving. The training was started with an initial learning rate of 5 x 10-4, which was reduced by a factor of 2 every time the validation loss became stagnant for 10 epochs. To prevent the network from overfitting, the data were augmented using mirroring along the vertical axis with a probability of 0.5.

\subsection{Radiomic feature extraction} \label{supp_methods_radiomics}
Radiomics features were extracted using the PyRadiomics 3.0.1 (\url{https://github.com/Radiomics/pyradiomics})~\cite{van2017computational} tool. This method produces a set of quantitative features for various combinations of input image and segmentation, as specified by the user. Based on our proposed framework, input images can include at most four scans belonging to the four segmentable sequences i.e., T1WI, Gd-T1WI, T2WI and FLAIR. As per our segmentation module discussed in Section~\ref*{sec:seg}, the input segmentations can include at most five different tumor classes viz. ED, NC, ET, TC, and WT. The default set of radiomics features were extracted including 3D shape features (n = 14), first order features (n = 18), and texture features (n = 75). Shape features are independent of image contrast and were therefore only extracted once per segmentation per session (5 segmentations × 14 features = 70 shape features per session). First order and texture features were extracted from each combination of image and segmentation (4 image contrasts × 5 segmentations × 93 first order and texture features = 1,860 features per patient). Thus, the complete radiomics feature-set included 1,930 individual image features (70 shape features + 1860 first order and texture features) per session. For all input images, intensity-normalized scans were used as described in Section~\ref*{sec:curation}. The parameter values for all feature extraction were set as default for ease of use and reproducibility.

\section{Supplementary Results}
\subsection{Segmentation results on BraTS 2021 dataset} \label{supp_results_segmentation}
\begin{figure}[]
\centering
\includegraphics[width=\textwidth]{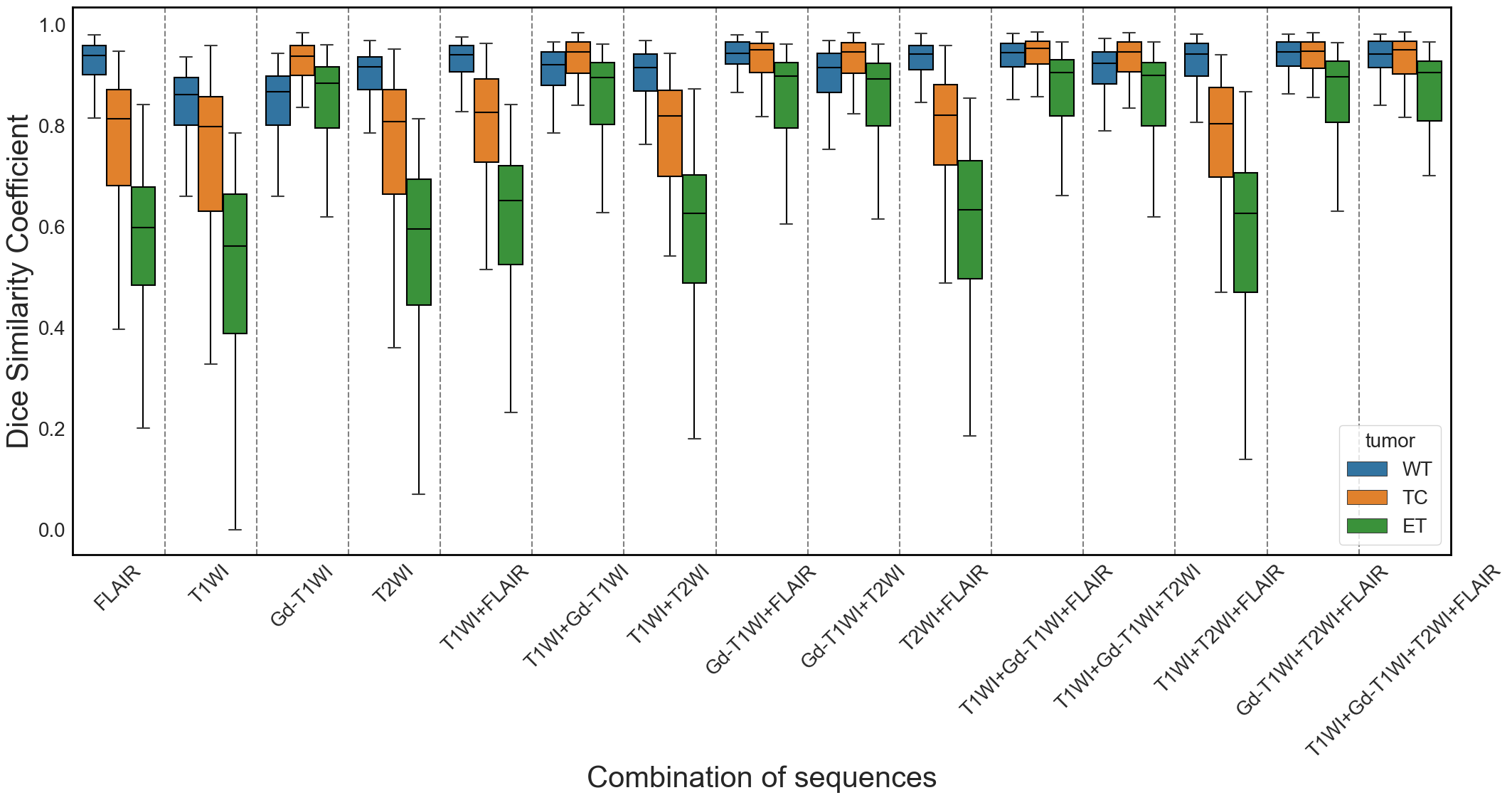}
\caption{Dispersion of Dice Similarity Coefficient (DSC) values, stratified by presence of sequences, showing the agreement in terms of overlap between the tumor segmentation masks predicted by convolutional neural network and the expert-annotated masks. WT = whole tumor, TC = tumor core, ET = enhancing tumor.} 
\label{fig:fig3}
\end{figure}

For the BraTS 2021 dataset (Supplementary Figure~\ref*{fig:fig3}, Supplementary Table~\ref*{supp_table_seg_results}), in presence of Gd-T1WI, the mean DSC values were high for all tumor segmentation classes viz. WT (0.90±0.03), TC (0.88±0.01), and ET (0.81±0.01). In absence of Gd-T1WI, the mean DSC for ET (0.27 drop, $P < .001$) and TC (0.15 drop, $P < .001$) dropped significantly.  Hence, in absence of Gd-T1WI, the deployed models on WUSM and MDA were used only for WT segmentation.

\begin{table}[!htbp]
\caption{Demographic and clinical information of patient data acquired from Washington University School of Medicine (WUSM) and M.D. Anderson Cancer Center (MDA).}
\centering
\begin{tabular}{l|ll}
\hline
          & WUSM (n = 384) & MDA (n = 30) \\  \hline
Age       & 56 (44 - 66)   & 58 (44 - 65) \\  \hline
Sex       &                &              \\  
~~~Female    & 154 (40.1\%)   & 15 (50.0\%)  \\ 
~~~Male      & 230 (59.9\%)   & 15 (50.0\%)  \\ \hline
WHO Grade &                &              \\ 
~~~G2        & 65 (16.93\%)   & 5 (16.67\%)  \\ 
~~~G3        & 63 (16.41\%)   & 3 (10.0\%)   \\ 
~~~G4        & 256 (66.67\%)  & 22 (73.33\%) \\ 
\hline
\end{tabular}
\label{supp_table_dataset}
\end{table}

\begin{table}[!htbp]
\centering
\caption {Segmentation performance stratified by presence of sequence on the BraTS 2021 dataset in terms of Dice Similarity Co-efficient (DSC) for Whole tumor, Tumor core and Enhancing tumor classes. The table shows the DSC for all possible configurations of sequences being either absent~($\circ$) or present~($\bullet$), in order of FLAIR ($F$), T1WI ($T_1$), Gd-T1WI ($T_1c$), T2WI ($T_2$).}\label{tab:big_table} 
\newcolumntype{s}{p{3cm}}
\newcolumntype{L}{>{\columncolor[gray]{0.9}$}c<{$}}
\begin{tabular}{LLLL|sss}
\hline
\multicolumn{4}{c|}{Sequences} & \multicolumn{3}{c}{Tumor class}\\
\hline
\;F\;&\,T_1\,&T_1c&\,T_2\,&\enskip Whole Tumor  &Tumor Core   &Enhancing Tumor \\
\hline
\bullet   & \circ   & \circ   & \circ    & \enskip0.914 (0.091) & 0.733 (0.221) & 0.531 (0.22)  \\ 
\circ   & \bullet   & \circ   & \circ    & \enskip0.819 (0.132) & 0.713 (0.212) & 0.501 (0.221) \\ 
\circ   & \circ   & \bullet   & \circ    & \enskip0.811 (0.158) & 0.88 (0.173)  & 0.805 (0.219) \\ 
\circ   & \circ   & \circ   & \bullet    & \enskip0.879 (0.114) & 0.719 (0.232) & 0.525 (0.231) \\ 
\bullet   & \bullet   & \circ   & \circ    & \enskip0.916 (0.085) & 0.757 (0.21)  & 0.567 (0.229)    \\ 
\circ   & \bullet   & \bullet   & \circ    & \enskip0.893 (0.103) & 0.868 (0.218) & 0.799 (0.242)    \\ 
\circ   & \bullet   & \circ   & \bullet    & \enskip0.885 (0.115) & 0.73 (0.235)  & 0.546 (0.23)     \\ 
\bullet   & \circ   & \bullet   & \circ    & \enskip0.926 (0.063) & 0.886 (0.175) & 0.818 (0.216)    \\ 
\circ   & \circ   & \bullet   & \bullet    & \enskip0.886 (0.111) & 0.875 (0.195) & 0.804 (0.232)    \\ 
\bullet   & \circ   & \circ   & \bullet    & \enskip0.924 (0.062) & 0.751 (0.209) & 0.563 (0.233)    \\ 
\bullet   & \bullet   & \bullet   & \circ    & \enskip0.924 (0.066) & 0.892 (0.175) & 0.825 (0.217) \\ 
\circ   & \bullet   & \bullet   & \bullet    & \enskip0.892 (0.121) & 0.869 (0.215) & 0.803 (0.247) \\ 
\bullet   & \bullet   & \circ   & \bullet    & \enskip0.916 (0.1)   & 0.74 (0.217)  & 0.554 (0.234) \\ 
\bullet   & \circ   & \bullet   & \bullet    & \enskip0.926 (0.071) & 0.882 (0.181) & 0.813 (0.223) \\ 
\bullet   & \bullet   & \bullet   & \bullet    & \enskip0.923 (0.07)  & 0.89 (0.169)  & 0.814 (0.224)    \\ 
\hline
\end{tabular}
\label{supp_table_seg_results}
\end{table}
\end{document}